\documentclass[12pt]{article}
\usepackage[T1]{fontenc}
\usepackage{amsfonts}
\usepackage{latexsym}
\usepackage{dsfont}
\usepackage{amsmath}
\usepackage{bbm}
\usepackage{amssymb}
\usepackage{amsthm}
\usepackage{mathrsfs}
\usepackage[nosort]{cite}
\usepackage{setspace}
\usepackage{color}
\usepackage{graphicx}
\usepackage{accents}
\usepackage[font=footnotesize,labelfont=bf]{caption}
\usepackage[utf8]{inputenc}
\usepackage{geometry}
\usepackage[toc,page]{appendix}

\usepackage{tikz}
\usetikzlibrary{arrows.meta,backgrounds,calc,fit,positioning,scopes,shadows}

\makeatletter
\def\tikzsavelastnodename#1{\let#1=\tikz@last@fig@name}
\makeatother

\usepackage[hidelinks]{hyperref}
\hypersetup{
colorlinks=false,
citecolor= blue,
linkcolor= blue,
urlcolor= blue,
breaklinks=true
}
\usepackage{cleveref}
\crefformat{section}{\S#2#1#3} % see manual of cleveref, section 8.2.1
\crefformat{subsection}{\S#2#1#3}
\crefformat{subsubsection}{\S#2#1#3}

\renewcommand{\theequation}{\thesection.\arabic{equation}}

\def\appendix#1{%
\setcounter{section}{0}%
\renewcommand{\thesection}{\Alph{section}}%
\refstepcounter{section}%
\setcounter{equation}{0}%
\section*{Appendix \thesection\protect\indent
\parbox[t]{11.15cm}{#1}}%
\addcontentsline{toc}{section}{Appendix \thesection\ \ \ #1}}

\allowdisplaybreaks
\numberwithin{equation}{section}

\makeatletter
 \let\old@startsection=\@startsection
 \let\oldl@section=\l@section
 \renewcommand{\@startsection}[6]{\old@startsection{#1}{#2}{#3}{#4}{#5}{#6\mathversion{bold}}}
 \renewcommand{\l@section}[2]{\oldl@section{\mathversion{bold}#1}{#2}}
\makeatother

\def\dd{\text{d}}

\makeatletter
\def\user@resume{resume}
\def\user@intermezzo{intermezzo}
\newcounter{previousequation}
\newcounter{lastsubequation}
\newcounter{savedparentequation}
\renewenvironment{subequations}[1][]{%
      \def\user@decides{#1}%
      \setcounter{previousequation}{\value{equation}}%
      \ifx\user@decides\user@resume 
           \setcounter{equation}{\value{savedparentequation}}%
      \else  
      \ifx\user@decides\user@intermezzo
           \refstepcounter{equation}%
      \else
           \setcounter{lastsubequation}{0}%
           \refstepcounter{equation}%
      \fi\fi
      \protected@edef\theHparentequation{%
          \@ifundefined {theHequation}\theequation \theHequation}%
      \protected@edef\theparentequation{\theequation}%
      \setcounter{parentequation}{\value{equation}}%
      \ifx\user@decides\user@resume 
           \setcounter{equation}{\value{lastsubequation}}%
         \else
           \setcounter{equation}{0}%
      \fi
      \def\theequation  {\theparentequation  \alph{equation}}%
      \def\theHequation {\theHparentequation \alph{equation}}%
      \ignorespaces
}{%
  \ifx\user@decides\user@resume
       \setcounter{lastsubequation}{\value{equation}}%
       \setcounter{equation}{\value{previousequation}}%
  \else
  \ifx\user@decides\user@intermezzo
       \setcounter{equation}{\value{parentequation}}%
  \else
       \setcounter{lastsubequation}{\value{equation}}%
       \setcounter{savedparentequation}{\value{parentequation}}%
       \setcounter{equation}{\value{parentequation}}%
  \fi\fi
  \ignorespacesafterend
}
\makeatother

\begin{document}

%%%%%%%%%%%%% TITLE %%%%%%%%%%%%%%%%%%%%%%%%%

\begin{titlepage}
\vspace*{-1.0cm}

\begin{center}
%\today

\vspace{2.0cm}

{\LARGE  {\fontfamily{lmodern}\selectfont \bf The all-loop spectrum of GGK strings
}}
\normalsize

\vskip 1.5cm
\textsc{Marius de Leeuw$^a$, Andrea Fontanella$^{a,b}$, Cameron Gibson$^c$ \footnotesize \\
and \normalsize Juan Miguel Nieto Garc\'ia$^d$}\\
\vskip 1.2cm

\begin{small}
{}$^{a}$ \textit{School of Mathematics $\&$ Hamilton Mathematics Institute, \\
Trinity College Dublin, Ireland}\\
\vspace{1mm}
\href{mailto:mdeleeuw[at]maths.tcd.ie}
{\texttt{mdeleeuw[at]maths.tcd.ie}}

\vspace{5mm}
{}$^{b}$ \textit{School of Mathematical and Statistical Sciences, \\ 
University of Galway, University Road, Galway H91 TK33, Ireland} \\
\vspace{1mm}
\href{mailto:andrea.fontanella[at]universityofgalway.ie}{\texttt{andrea.fontanella[at]universityofgalway.ie}}

\vspace{5mm}
{}$^{c}$ \textit{Mitchell Institute for Fundamental Physics \& Astronomy,\\
         Texas A\&M University,
         College Station,
         TX 77843-4242,
         USA.} \\
\vspace{1mm}
\href{mailto:camerongibson[at]tamu.edu}
{\texttt{camerongibson[at]tamu.edu}}

\vspace{5mm}
{}$^{d}$ \textit{Departamento de Matem\'atica Aplicada a las TIC,\\
ETSIS de Telecomunicaci\'on, Universidad Polit\'ecnica de Madrid \\
C. Nikola Tesla s/n, $28031$ Madrid, Spain} \\
\vspace{1mm}
\href{mailto:juanmiguel.nietogarcia[at]upm.es}{\texttt{juanmiguel.nietogarcia[at]upm.es}}

\end{small}

\end{center}

\vskip 0.8 cm
\begin{abstract}
\vskip0.8cm

\noindent We study the non-perturbative spectrum of the non-relativistic limit of string theory in AdS$_5\times$S$^5$, also known as GGK theory. In light-cone gauge, we show that the bosonic spectrum on a number of vacua is described at all-loops by free massive and massless scalar fields in AdS$_2$, confirming a conjecture by de Leeuw, Fontanella and Nieto-García, \href{https://arxiv.org/abs/2403.09563}{arXiv:2403.09563}. In static gauge, we obtain the same result for any classical string vacuum, and including fermions. We compute the quantum correction to the classical energy, and show that it vanishes once we adopt the regularised Hurwitz series. We conclude that the quantum energy spectrum built upon any GGK string consists solely of its classical vacuum energy.

\end{abstract}

\end{titlepage}

\tableofcontents
\vspace{5mm}
\hrule

%%%%%%%%%%%%%%%%%%% BODY %%%%%%%%%%%%%%%%%%

\setcounter{section}{0}
\setcounter{footnote}{0}

\section*{Introduction}

The AdS/CFT correspondence is a concrete realisation of the holographic principle, namely the idea that a gravity and a gauge theory are nothing else than two sides of the same coin \cite{Maldacena:1997re,Gubser:1998bc,Witten:1998qj}. Although being a very powerful statement, this idea comes with its limitations. 
In particular, holography is currently best understood when the gravitational theory is defined on an asymptotically anti-de Sitter spacetime.
Recently, new ideas regarding limits of the AdS/CFT correspondence have been proposed. By limit, we mean identifying a physical parameter of the theory and taking it to a critical value. This top-down approach allows to: i) derive a new example of holography from  AdS/CFT, and ii) zoom into a corner of AdS/CFT, sometimes more tractable. Limits of AdS/CFT have been explored from different point of views: the non-relativistic limit (i.e. infinite speed of light) \cite{Fontanella:2024rvn, Fontanella:2024kyl, Lambert:2024uue, Lambert:2024yjk, Lambert:2024ncn, Joe_thesis, Kim:2026hpe, Blair:2024aqz, Blair:2025prd, Harmark:2025ikv, Guijosa:2025mwh, Harmark:2017rpg, Harmark:2018cdl, Harmark:2020vll}, the Carroll limit (i.e. vanishing speed of light) \cite{Fontanella:2025tbs}, and the flat space limit (i.e. infinite AdS radius) \cite{Fontanella:2025tbs, Eden:2026ulx, Lipstein:2025jfj, Alday:2024yyj, Berenstein:2025tts, Berenstein:2025qhb, Hijano:2019qmi, Ciambelli:2018wre, Campoleoni:2023fug, Li:2021snj, Hijano:2020szl, Kulkarni:2025qcx, Kraus:2024gso, Duary:2022pyv, Fareghbal:2024lqa, Duary:2024kxl, Marotta:2024sce, Surubaru:2025fmg, Hao:2025btl, Arenas-Henriquez:2025rpt, Poulias:2025eck, Neuenfeld:2025wnl, Grumiller:2023rzn, Bagchi:2023fbj}. For recent reviews on the topics, see \cite{Fontanella:2026gaq, Fontanella:2026fzs, Ruzziconi:2026bix, Nguyen:2025zhg, Bagchi:2025vri}.     

In this article, we will focus on the non-relativistic limit of AdS$_5$/CFT$_4$. In particular, we shall focus on the proposed holographic duality \cite{Fontanella:2024rvn, Fontanella:2024kyl} between the Gomis-Gomis-Kamimura (GGK) non-relativistic limit of Type IIB superstring in AdS$_5\times$S$^5$ \cite{Gomis:2005pg}\footnote{GGK theory is an example of non-relativistic string theory where the target space is a String Newton-Cartan manifold and the worldsheet is relativistic \cite{Andringa:2012uz, Bergshoeff:2015uaa, Bergshoeff:2018yvt, Bergshoeff:2019pij}. The flat space analogue of it is the Gomis-Ooguri string \cite{Gomis:2000bd, Danielsson:2000gi}. For reviews on the topic, see \cite{Oling:2022fft} and also \cite{Fontanella:2026gaq}.} and the Galilean Yang-Mills theory in 3+1 dimensions with five interacting scalar fields, often abbreviated as \emph{GGK/GYM duality} (see \cite{Fontanella:2026gaq, Fontanella:2026fzs} for reviews). This holographic duality was proposed as the result of a consistent limit of AdS/CFT, where i) the non-relativistic and the $\alpha'\to 0$ limits commute; ii) the stack of D3-branes retains a horizon region in the non-relativistic limit; iii) the GYM with scalars theory is defined on the Penrose boundary of GGK string theory; and iv) the symmetries of the two theories match, see also \cite{Fontanella:2024hgv}. An important test that is still missing is a quantitative test, showing that observables from the two theories can be mapped into each others with matching expectation values.

In view of a quantitative test of the GGK/GYM duality, in this article we consider the problem of finding the energy spectrum of the GGK string. This problem has been tackled in the past from different angles. A family of classical string solutions, acting as vacua for the quantisation of the theory, were found in \cite{Fontanella:2021btt}. The simplest classical solution is a static string, which has zero energy. Quantisation and string spectrum around this solution was computed in \cite{Sakaguchi:2007ba}. However, the fact that the classical energy of the static string is zero renders this vacuum unattractive from the holographic program point of view, since it would potentially correspond to dual operators with vanishing classical scaling dimensions.       

The simplest vacuum with a holographic application is a folded BMN-like string\footnote{The name should not confuse the reader with the well-known BMN string. The BMN string is a point-like solution, which is ``too fast'' to be seen by the GGK theory. The BMN string is not part of GGK theory's spectrum \cite{Fontanella:2023men}. }, which has a classical dispersion relation of the type $E \sim J^2$, with $J$ being a linear momentum along a flat $\mathbb{R}^5$ direction. Quantisation around this vacuum was first discussed in \cite{deLeeuw:2024uaq}, following the light-cone gauge fixing in \cite{Fontanella:2021hcb}. Expanding perturbatively around the folded BMN-like solution in light-cone gauge, it was demonstrated that the bosonic sector of the GGK theory, up to and including terms of sixth order in the fields, reduces to an action describing three massive and five massless free scalar fields in AdS$_2$ \cite{deLeeuw:2024uaq}. In the same article, the authors suggested a route how to extend the result in a fully non-perturbative fashion, which amounts to solving a complicated PDE. The purpose of the present article is to solve this problem using a different method.

\paragraph{Plan of this article.}  This article is divided into two parts. In the first part (Section \ref{sec:LC_gauge}), we consider the spectral problem of the GGK theory in the \emph{light-cone gauge}. We propose a method that captures the full non-perturbative expansion of the GGK action around the folded BMN-like string based on a redefinition of the spatial worldsheet coordinate. Then, we extend the method to other more involved classical string solutions, and show that the bosonic sector of the GGK action maps into a theory of three massive and five massless free scalar fields in AdS$_2$ also in all these cases.

In the second part (Section \ref{sec:static_gauge}), we consider the spectrum in the \emph{static gauge}. In this gauge, computations become extremely simple, and we show that the full GGK action (including fermions) expands around a \emph{generic closed string classical solution} as three massive and five massless free scalar fields and eight massive Majorana 2d free fermions in AdS$_2$. This result is exact at all loops, and in agreement with the result of Section \ref{sec:LC_gauge}.\footnote{The agreement holds for the bosonic sector, so far. Fermions in the light-cone gauge will not be considered in this analysis, because technically more complicated to deal with.}  This is a good cross check of our results, since physical quantities cannot be gauge-dependent. Finally, we compute the non-perturbative spectrum of the GGK string, and we show that there is an exact cancellation between the bosonic and fermionic contributions, once the Hurwitz series regularization is adopted. We find that the full quantum spectrum of the GGK string is only made of the classical vacuum energy.

\section{Spectrum in light-cone gauge}
\label{sec:LC_gauge}

\subsection{Short review of \cite{deLeeuw:2024uaq}}

We are considering the non-relativistic limit of the bosonic sector of Type IIB string theory in AdS$_5\times S^5$. We use Cartesian coordinates $(t,z_i)$ for AdS$_5$ and $(\phi,y_i)$ for $S^5$, with $i=1,...,4$. We further split $z_i\equiv (z_1,z_m)$ with $m=2,3,4$, such that the non-relativistic limit scales $t$ and $z_1$ as longitudinal coordinates. We denote string worldsheet coordinates by $\sigma^\alpha=(\tau,\sigma)$ with $\sigma \simeq \sigma+2\pi$. We take $X^\mu(\tau,\sigma) = (t,z_i, \phi, y_i )$ to be the string embedding fields. We denote $\dot{X}\equiv \partial_\tau X$ and $X' \equiv \partial_\sigma X$, or equivalently, $X_{\tau} \equiv \partial_\tau X$ and $X_{\sigma} \equiv \partial_\sigma X$. We also denote the string tension by $T$. The light-cone gauge fixing procedure eliminates $t$ and $\phi$, giving us an action for the fields transverse to the light-cone, $(z_i, y_i)$. For the full detail on the light-cone gauge fixing procedure, see \cite{deLeeuw:2024uaq, Fontanella:2021hcb}, and for detail on the non-relativistic limit in Cartesian coordinates, see the review \cite{Fontanella:2026gaq}.

The non-relativistic action in light-cone gauge is then
\begin{eqnarray}
\label{gauge_fixed_action}
    S^{a=0}_{\text{g.f.}} &=& \int \dd^2 \sigma \, \frac{1}{8T (z_1^2 - 4)^3 (z_1^2 + 4) z_1^{\prime}} \bigg[ z_1^{10} (1 + T^2 y_j^{\prime 2}) - 4 z_1^8 (1 + T^2 y_j^{\prime 2} - 2 T z_1^{\prime}) \\
\notag
    &-& 16 z_1^6 \bigg(2 + 4 T z_1^{\prime} - T^2 z_m^{\prime 2} + T^2 z_1^{\prime 2} \dot{y}_j^2 - 2 T^2 y_j^{\prime} \dot{y}_j z_1^{\prime}  \dot{z}_1 + \dot{z}_1^2 + T^2 y_j^{\prime 2} (2 + \dot{z}_1^2)\bigg) \\
\notag
    &+& 32 z_1^4 \bigg( 4 - T^2 z_m^2 z_1^{\prime 2} + 2 T^2 z_m^{\prime 2} + 6 T^2 z_1^{\prime 2} \dot{y}_j^2  - 12 T^2 y_j^{\prime} \dot{y}_j z_1^{\prime}  \dot{z}_1 + 6 \dot{z}_1^2 + 2 T^2 y_j^{\prime 2} (2 + 3 \dot{z}_1^2) \bigg) \\
\notag
    &-& 256 z_1^2 \bigg(-1 + T^2 z_m^{\prime 2} - 6 T^2 y_j^{\prime} \dot{y}_j z_1^{\prime}  \dot{z}_1 + 3 \dot{z}_1^2 + T^2 y_j^{\prime 2} (-1 + 3 \dot{z}_1^2) - 2 T z_1^{\prime} (2 + T z_m^{\prime} \dot{z}_m \dot{z}_1 ) \\
\notag
    &+& T^2 z_m^{\prime 2} \dot{z}_1^2 + T^2 z_1^{\prime 2} (2 z_m^2 + 3 \dot{y}_j^2 + \dot{z}_m^2 ) \bigg) - 512 \bigg( 4T^2 y_j^{\prime} \dot{y}_j z_1^{\prime}  \dot{z}_1 - 2 T^2 y_j^{\prime 2} (-1 + \dot{z}_1^2) \\
\notag
    &-& 2(1 + T^2 z_m^{\prime 2})(-1 + \dot{z}_1^2) + 4T z_1^{\prime} (1 + T z_m^{\prime} \dot{z}_m \dot{z}_1 ) + T^2 z_1^{\prime 2} \big(3 z_m^2 - 2 (\dot{y}_j^2 + \dot{z}_m^2) \big) \bigg) \bigg]\, ,  
\end{eqnarray}
where sums over $m \in \{2, 3, 4\}$ and $j\in\{1,..., 4\}$ are to be understood.

To perform the quantisation of the action, we first choose a classical solution of the equations of motion around which we expand. The simplest solution we can use is the BMN-like string \cite{Fontanella:2021btt}, given by
\begin{equation}
    t=\tau \,, \qquad z_1=2\tan \left(-\frac{\kappa \sigma}{2}\right) \,, \qquad \phi = \nu \tau \,, \label{vacuum}
\end{equation}
with $\nu$ fixed by the choice of gauge, and with $\kappa \in \mathbb{Z}_{\neq 0}$.

As the classical theory corresponds to the large tension limit, the inverse of the tension is a convenient expansion parameter for the quantisation
\begin{equation}
    \hbar\equiv \frac{1}{\sqrt{T}} \, .
\end{equation}
Now we are ready to expand the transverse fields $X^I$ around the folded BMN-like vacuum,
\begin{eqnarray}
    z_1 = 2 \tan \left( -\frac{\kappa}{2} \sigma \right) + \hbar \, \delta z_1 \, , \qquad     z_m =  \hbar \, \delta z_m \, , \qquad     y_j = \hbar \, \delta y_j \, .
\end{eqnarray}
where $\delta z_1, \delta z_m, \delta y_j$ are the fluctuation fields, and they must satisfy periodic boundary conditions. The gauge fixed action expands in small $\hbar$ as follows\footnote{To facilitate the computations, it is useful to eliminate the $\hbar$ dependence that appears in several places inside the gauge fixed action \eqref{gauge_fixed_action}, before taking the expansion \eqref{exapnsion_BMN_vacuum}. This is achieved in the same way as it was done for the relativistic action, that is, by rescaling $\sigma \to T\sigma$.}
\begin{equation}\label{exapnsion_BMN_vacuum}
   S^{a=0}_{\text{g.f.}}=\int \dd^2 \sigma \sum _{j=0}^\infty \hbar^{j-2}\mathcal{L}_j \,,
\end{equation}
where $\mathcal{L}_i$ are homogeneous polynomials of degree $i$ in the field fluctuations and their derivatives. $\mathcal{L}_0$ is a constant, $\mathcal{L}_1$ is a total derivative and $\mathcal{L}_2$ can be transformed by a simple redefinition
\begin{equation}
    \begin{split}
        &\delta z_1\rightarrow \sqrt{2} \kappa^\frac{3}{2}\sec^2 \left(\frac{\kappa \sigma}{2}\right) \delta z_1 \,,\\
        &\delta z_m \rightarrow \frac{\sqrt{2\kappa}}{\cos^2\left(\frac{\kappa \sigma}{2}\right)\sec(\kappa \sigma)}\delta z_m \,,\\
        &\delta y_j \rightarrow \sqrt{2\kappa} \delta y_j \,, \qquad \tau \rightarrow \kappa \tau \,.
    \end{split}
\end{equation}
These transformations turn $\mathcal{L}_2$ into the free-field action for 3 massive fields $\delta z_m$ and 5 massless fields $(\delta z_1,\delta y_j)$ in AdS$_2$.  Surprisingly, it was shown in \cite{deLeeuw:2024uaq} that interaction terms including up to sixth order vanish after further perturbative field redefinitions, i.e. $\mathcal{L}_3=\mathcal{L}_4=\mathcal{L}_5=\mathcal{L}_6=0$ (up to boundary terms). The field redefinitions took quite a complicated form, motivating a search for a nonlinear field redefinition to prove equivalence to the free-field action to all orders. We present such a redefinition in the following subsection.

\subsection{The idea applied to the folded BMN-like string}

In this section we present a method that is capable of capturing the all-loop perturbation theory around the vacuum \eqref{vacuum}. We start by introducing a new worldsheet variable $\rho$, defined as
\begin{equation}
\sigma=\rho+ \frac{\hbar}{\sqrt{2\kappa}} X(\tau,\rho) \, . \label{rhotosigma}
\end{equation}
The idea is to use the classical string solution expressed in the $\rho$ variable as the non-perturbative field redefinition that was sought in \cite{deLeeuw:2024uaq}. This amounts to  
\begin{equation}
\begin{split}
    &z_1(\tau,\sigma(\tau,\rho))=2\tan\left(-\frac{\kappa \rho}{2}\right)\,,\\
    &z_m(\tau,\sigma(\tau,\rho))=\frac{\hbar}{\sqrt{2\kappa}}\cos(\kappa \rho)\sec^2\left(\frac{\kappa\rho}{2}\right)Z_m(\tau,\rho)\,,\\
    &y_i(\tau,\sigma(\tau,\rho))=\frac{\hbar}{\sqrt{2\kappa}} Y_i(\tau,\rho)\, . \label{fieldsredefs}
    \end{split}
\end{equation}
Intuitively, $X(\tau, \rho)$ in \eqref{rhotosigma} will play the role of a fluctuation field for $z_1$, whereas $Z_m(\tau, \rho)$ and $Y_i (\tau, \rho)$ will be the fluctuations for $z_m$ and $y_i$ respectively.  

By substituting these new variables, and changing the integral over $\sigma$ to an integral over $\rho$, the action \eqref{gauge_fixed_action} becomes
\begin{align}
    S_{\text{f}} &=\int \dd\tau \dd \rho \,  \left[  \frac{1}{4}\left(\hbar^4\left(X_{\tau}^2-\frac{1}{\kappa^2}X_{\rho}^2\right)+Y_{j,\tau}^2-\frac{1}{\kappa^2}Y_{j,\rho}^2+Z_{m,\tau}^2-\frac{1}{\kappa^2}Z_{m,\rho}^2\right) -\frac{1}{2}Z_m^2\sec^2(\kappa \rho)\right. \notag \\
    &\left.+1-\frac{\hbar^2}{2\kappa}+\partial_\rho\left(\frac{1}{4\kappa}Z_m^2\sec(\kappa\rho)\tan\left(\frac{\kappa \rho}{2}\right)+\left(\frac{\hbar}{\sqrt{2\kappa}}-\frac{\hbar^3}{\sqrt{2}\kappa^\frac{3}{2}}\right)X\right)\right]\, . \label{bmn}
\end{align}
For a detailed derivation, we refer to Appendix~\ref{deriv}. Notice that the second line is made of either constants or boundary terms, so they do not contribute to the equations of motion of the fluctuation fields. Therefore, $X$ and $Y_i$ are five free massless fields and $Z_m$ are three free massive fields in AdS$_2$. The $\hbar^4$ factor in front of the kinetic term of $X$ can be absorbed by a rescaling $\hbar^2 X \to X$, since we have not performed any expansion in $\hbar$ yet.

As we are interested in closed strings, the range of $\sigma$ is between $0$ and $2\pi$. In contrast, the range of $\rho$ is not a priori fixed, since it depends on the field $X(\tau, \rho)$. If we demand that $X(\tau, \rho)$ is periodic in $\rho$ with period $2\pi$, the range of integration in $\rho$ takes the form $[\upsilon(\tau), \upsilon(\tau) + 2\pi]$ for some function $\upsilon(\tau)$ that satisfies
\begin{eqnarray}
   0 = \upsilon(\tau) + \frac{\hbar}{\sqrt{2\kappa}} X(\tau, \upsilon(\tau)) \, .  
\end{eqnarray}
The range of $\rho$ is also $2\pi$, but with a starting point that moves accordingly to the function $\upsilon(\tau)$. As we will impose $2\pi$-periodicity for $Y_i$ and $Z_m$, the Lagrangian will be $2\pi$ periodic. As a consequence, the integral from $0$ to $\upsilon (\tau)$ will have the same values as the integral from $2\pi$ to $2\pi + \upsilon(\tau)$, allowing us to change the range of integration to the usual $[0,2\pi]$.

\subsection{Recovering the perturbative description}

In this section, we motivate the non-perturbative method presented earlier by making contact with the perturbative analysis of \cite{deLeeuw:2024uaq}.

The first step is to invert \eqref{rhotosigma} and write $\rho$ as a function of $\sigma$. This is not straightforward, as the field $X$ depends on $\rho$. 
First, we notice that we can rewrite \eqref{rhotosigma}, for fixed $\tau$, as 
\begin{equation}
    \rho=\sigma-\epsilon X(\rho)\, , \qquad\qquad \epsilon \equiv \frac{\hbar}{\sqrt{2\kappa}} \, .
\end{equation}
We can apply the Lagrange's inversion theorem to this formula, and write any function of $\rho$ as an infinite sum
\begin{equation}\label{Lagrange}
    f(\rho(\sigma))=f(\sigma)+\sum_{n=1}^\infty \frac{(-\epsilon)^n}{n!}\partial_\sigma^{n-1}\left(X^n(\sigma)f'(\sigma)\right)  \,.
\end{equation}
Applying this formula to Eqs.~\eqref{fieldsredefs} gives
\begin{align}
    z_1=&-2\tan\left(\frac{\kappa \sigma}{2}\right)+\epsilon\kappa \sec^2\left(\frac{\kappa \sigma}{2}\right)\left[X-\frac{1}{2} X\left(\kappa \tan\left(\frac{\kappa\sigma}{2}\right)X+2X'\right)\epsilon\right.\\
    &\left.+\epsilon^2 \left(\frac{1}{2}X^2X''+\kappa X^2X'\tan\left(\frac{\kappa \sigma}{2}\right)+XX'^2+\frac{1}{12}\kappa^2X^3\left(3\sec^2\left(\frac{\kappa \sigma}{2}\right)-2\right)\right)+\mathcal{O}(\epsilon^3)\right] \,,  \notag \\
    z_m=&\hphantom{-} \epsilon\cos(\kappa \sigma)\sec^2\left(\frac{\kappa \sigma}{2}\right)\left[Z_m+\epsilon X \left(\kappa \sec(\kappa \sigma)\tan\left(\frac{ \kappa \sigma }{2}\right)Z_m-Z'_m\right)\right. \notag \\
    &+\epsilon^2\frac{1}{4}X\left[4X'\left(Z_m'-\kappa Z_m \sec(\kappa \sigma)\tan\left(\frac{\kappa \sigma}{2}\right)\right)\right.\\
    &\left.+\left.X\left(2Z_m''+\kappa \sec^2\left(\frac{\kappa \sigma}{2}\right)((\kappa-2\kappa \sec(\kappa \sigma))Z_m-2Z_m'\tan(\kappa \sigma))\right)\right]+\mathcal{O}(\epsilon^3)\right] \,, \notag \\
    y_i=&\hphantom{-} \epsilon\left[Y_i-\epsilon X Y_i'+\frac{\epsilon^2}{2}(X^2Y_i''+2XX'Y_i')+\mathcal{O}(\epsilon^3)\right]\,. \label{eqstomap}
\end{align}
where all the functions should be understood as functions $(\tau,\sigma)$ and $'$ denotes a derivative with respect to $\sigma$. After applying the map
\begin{equation}
    (X,Z_m,Y_i)=2\kappa (\delta z_1,\delta z_m, \delta y_i) \, ,
\end{equation}
we recover the field redefinition computed in \cite{deLeeuw:2024uaq}. This works as a check for the computations presented above.

\subsection{Other vacua}

\paragraph{The strategy.} The procedure presented in the previous sections is not limited to the folded BMN-like solution and can be applied to all the classical solutions found in \cite{Fontanella:2021btt}. When considering more generic classical solutions, there is an apparent obstacle in using the method: the redefinition from $\sigma$ to $\rho$ is crafted to absorb the fluctuations around the coordinate with the non-trivial profile, $z_1$. Thus, it is not clear how to define $\rho$ if we have several directions with a non-trivial profile. This is actually not that difficult. If we follow the recipe blindly until we arrive to the action \eqref{bmn}, we see that the action is quadratic in all the fields and they are all decoupled. At this point, if $Y_i$ or $Z_m$ gain a non-trivial classical solution, $Y_i^{\text{cl}}$ and $Z_m^{\text{cl}}$, then we only have to perform the following shift transformation
\begin{equation}
    Y_i\to Y_i^{\text{cl}}+\delta Y_i \,, \qquad Z_m\to Z_m^{\text{cl}}+\delta Z_m \,. 
\end{equation}
As the action is quadratic in the fields, the transformed action can be written as
\begin{align}
    S(Y_i^{\text{cl}}+\delta Y_i , Z_m^{\text{cl}}+\delta Z_m)&=S(Y_i^{\text{cl}}, Z_m^{\text{cl}}) +S(\delta Y_i, \delta Z_m) + (\text{boundary terms}) \,.
\end{align}
This happens because, when we expand the functional, the linear terms have the form
\begin{equation}
    \int{ \dd\tau \dd\sigma \left.\frac{\delta S (Y_k, Z_m)}{\delta Y_i(\tau, \sigma) }\right|_{\substack{Y_k\to Y_k^{\text{cl}} \\Z_m \to Z_m^{\text{cl}}}} \delta Y_i (\tau, \sigma)} \, ,
\end{equation}
and similarly for $Z_m$, which vanish because the functional derivative is nothing but an Euler-Lagrange equation evaluated on a solution. Furthermore, the quadratic terms have the form
\begin{equation}
    \int{ \dd\tau_1 \dd\sigma_1 \dd\tau_2 \dd\sigma_2 \left.\frac{\delta^2 S (Y_k, Z_m)}{\delta Y_i(\tau_1, \sigma_1) \delta Y_j(\tau_2, \sigma_2)}\right|_{\substack{Y_k\to Y_k^{\text{cl}} \\Z_m \to Z_m^{\text{cl}}}} \delta Y_i (\tau_1, \sigma_1) \delta Y_j (\tau_2, \sigma_2)}=S(\delta Y_i, 0) \,,
\end{equation}
and similarly for $Z_m$, because the action is quadratic and separable in fields. These two properties of the action also imply that $S(\delta Y_i, \delta Z_m)=S(\delta Y_i, 0)+S(0, \delta Z_m)$.

In the rest of this section, we will show this mechanism in action for some concrete classical solutions.\footnote{In principle, subtleties could arise from poles within, for example, the $\sec(\kappa \rho)$ term. However, in all the cases we consider, the standard regularisation can be used.}

\paragraph{Computing classical solutions.} Consider the Euler-Lagrange equations of the transformed Lagrangian \eqref{bmn}. $X$ and $Y_i$ are simply free massless fields, so solving their equations of motion is trivial. The equations of motion for each of the $Z_m$ are more involved and take the form
\begin{equation}
Z_{m,\tau \tau }-\frac{1}{\kappa^2}Z_{m,\rho \rho}+2Z_m\sec^2(\kappa \rho)=0 \label{pt} \,.
\end{equation}
This is a known differential equation, an $l=1$ P\"oschl-Teller potential, and can be solved exactly (see, for example, \cite{Dutt_2001}). In particular, as we are interested in closed strings, $Z_m(\tau,\rho)=Z_m(\tau,\rho+2\pi)$, we can perform the following transformation
\begin{equation}
    Z_m(\tau,\rho)=\left(\frac{1}{\kappa}\partial_\rho+\tan(\kappa \rho)\right)G_m(\tau,\rho) \,, \label{gsol}
\end{equation}
where the functions $G_m$ satisfy a wave equation with $\kappa^{-1}$ as the speed of sound
\begin{equation}
G_{m,\tau\tau}-\frac{1}{\kappa^2}G_{m,\rho\rho}=0\,. \label{wave}
\end{equation}
The most generic solution to a wave equation is the sum of left-moving and right-moving waves
\begin{equation}
G_m=G_{m,+}(\tau+\kappa\rho)+G_{m,-}(\tau-\kappa\rho) \,,
\end{equation}
for arbitrary functions $G_+$ and $G_-$. Expanding these functions in Fourier modes thanks to the periodicity in $\rho$, we can write it as
\begin{equation}
\begin{split}
    G_m(\tau, \rho)=&\alpha_m+\beta_m \tau\\
    +&\sum_{n\in \mathbb{Z}}\left[A_{m,n}\cos\left(\frac{n}{\kappa}\tau\right)+B_{m,n}\sin\left(\frac{n}{\kappa}\tau\right)\right]\cos\left(n \rho\right)\\
    +&\sum_{n\in \mathbb{Z}}\left[C_{m,n}\cos\left(\frac{n}{\kappa}\tau\right)+D_{m,n}\sin\left(\frac{n}{\kappa}\tau\right)\right]\sin\left(n \rho\right) \,.
\end{split} \label{gmodes}
\end{equation}
We can now write a general class of solutions to the equation of motion for $Z_m$ by applying the operator \eqref{gsol} to this expansion.

\paragraph{Generalised folded BMN-like string.}
The generalised folded BMN-like solution with closed boundary conditions is given by 
\[
 G^{\text{cl}}_m(\rho)=-{ \sqrt{\kappa}\over \sqrt{2}\,\hbar}\,C^1_m \,, \qquad Z^{\text{cl}}_m(\rho)=-{ \sqrt{\kappa}\over \sqrt{2}\,\hbar}\,C^1_m\tan(\kappa\rho) \, .
\]
Then, under the map $Z_m\to Z_m^{\text{cl}}+\delta Z_m$, we get the Lagrangian
\begin{equation}
    \mathcal{L}_{\text{gf}}=\mathcal{L}_\text{f}+\partial_\rho\left(-\frac{ {\left(C_m^1\right)}^2}{8\hbar^2}\sec(\kappa \rho)\tan(\kappa \rho)+\frac{\sqrt{2}}{4\hbar\sqrt{\kappa}}\sec({\kappa\rho})C_m^1Z_m\right) \,.
\end{equation}
where $\mathcal{L}_\text{f}$ is the Lagrangian expanded around the folded BMN-like string \eqref{bmn}. As the additional terms are total derivatives, the action becomes again free fields in AdS$_2$. 

Note that under the change of variables in \eqref{fieldsredefs}, this string solution is equivalent to
\begin{equation}
      z^{\text{cl}}_m=C_m^1\tan\left(-\frac{\kappa \rho}{2}\right) \,,
\end{equation}
which is the generalised folded BMN-like string with closed boundary conditions from \cite{Fontanella:2021btt}.

\paragraph{GKP-like string.}
A possible solution to Eq.~\eqref{wave} would be
\begin{equation}
    G(\tau,\rho)=g(\rho) \big(\cos(\omega \tau),\sin(\omega\tau),0\big) \,,
\end{equation}
with
\begin{equation}
    g''(\rho)+\kappa^2\omega^2g(\rho)=0 \, .
\end{equation}
Using the map \eqref{gsol}, we can write a solution for the equations of motion of $Z_m$
\begin{equation}
    Z^{\text{cl}}_m(\tau,\rho)=f(\rho)\big(\cos(\omega \tau),\sin(\omega\tau),0\big)\,,
\end{equation}
where $f(\rho)$ is defined as 
\begin{equation}
    f(\rho)=\frac{1}{\kappa}g'(\rho)+\tan(\kappa\rho)g(\rho) \,,
\end{equation}
and thus fulfils the differential equation
\begin{equation}
    f''(\rho)+\kappa^2(\omega^2-2\sec^2(\kappa\rho))f(\rho)=0\,. \label{transf}
\end{equation}

Due to the rotation of the classical solution, rather than introducing the fluctuation fields straightforwardly, it is quicker to make the computations in co-moving cylindrical coordinates
\begin{equation}
\begin{split}
    (Z_2,Z_3,Z_4)\longrightarrow \ &(Z^{\text{cl}}_2,Z^{\text{cl}}_3,Z^{\text{cl}}_4)\\
    +&S(\tau,\rho)(\cos(\omega\tau),\sin(\omega\tau),0) \\
    +&P(\tau,\rho)(-\sin(\omega\tau),\cos(\omega\tau),0)\\
    +&R(\tau,\rho)(0,0,1) \,.
    \end{split}
\end{equation}
This will produce the following action
\begin{equation}
\begin{split}
\mathcal{L}_{\text{GKP}}
&=
\frac{1}{4}\left(\hbar^4\left(X_{\tau}^2-\frac{1}{\kappa^2}X_{\rho}^2\right)+Y_{j,\tau}^2-\frac{1}{\kappa^2}Y_{j,\rho}^2\right)\\
&+
\frac14\left[(S_\tau-\omega P)^2+(P_\tau+\omega S)^2+R_\tau^2-\frac1{\kappa^2}\left(S_\rho^2+P_\rho^2+R_\rho^2\right)\right]\\
&-\frac12\sec^2(\kappa\rho)\left(S^2+P^2+R^2\right)\\
&+\text{boundary terms} \,.
\end{split}
\end{equation}
The derivative expressions $(S_\tau-\omega P)$ and $(P_\tau+\omega S)$ are simply $\tau$-derivatives in the rotating frame. We can eliminate them by performing the redefinition $S+iP=e^{-i\omega \tau} (U+iV)$, that is, transforming back into fluctuations in Cartesian coordinates, giving us
\begin{align}
\mathcal{L}_{\text{GKP}}
&=
\frac{1}{4}\left(\hbar^4\left(X_{\tau}^2-\frac{1}{\kappa^2}X_{\rho}^2\right)+Y_{j,\tau}^2-\frac{1}{\kappa^2}Y_{j,\rho}^2\right)\\
&+
\frac14\left[U_\tau^2+V_\tau^2+R_\tau^2-\frac1{\kappa^2}\left(U_\rho^2+V_\rho^2+R_\rho^2\right)\right]-\frac12\sec^2(\kappa\rho)\left(U^2+V^2+R^2\right) \notag \\
&+\text{boundary terms} \,.\notag 
\end{align}
Once again, we obtain an action for free fields in AdS$_2$ also in this GKP-like background.

A particularly nice aspect of this formulation is that Eq.~\eqref{transf} here is exactly equivalent to the complicated differential equation (3.27) in \cite{Fontanella:2021btt} under the changes of variables
\begin{equation}
    x=-\tan\left(\frac{\kappa \rho}{2}\right)\,, \qquad z=(1-x^2)f \,, \qquad \omega \rightarrow \kappa \omega \,.
\end{equation}
These coordinates explain why the equation was solvable despite looking so complicated.

%%%%%%%%%%%%%%%%%%%%%%%%%%%%%%%%%%%%%  Static gauge %%%%%%%%%%%%%%%%%%%%%%%%%%%%%%%%%%%%%%%%%%

\section{Spectrum in static gauge}
\label{sec:static_gauge}

We now move to a different gauge. In this section, we show that we can perform the large string tension expansion of GGK theory, including fermions, in static gauge and around any vacua. Computations are simpler than in light-cone gauge, and the final results concerning the bosonic sector match in the two different gauges.    

\subsection{The bosonic sector}

We first focus on the bosonic sector. We are interested in the spectrum of the theory around closed string solutions. In order for the static gauge to be meaningful for closed string solutions, we need to describe the GGK theory in the so-called \emph{GGK coordinates} \cite{Gomis:2005pg, Fontanella:2026gaq}. In this set of coordinates, the Nambu-Goto GGK action is 
\begin{equation}\label{GGK_NG}
    S_{\text{GGK, bos}} = - T \int \dd^2 \sigma \sqrt{|\tau|} \tau^{\alpha\beta} H_{\mu\nu} \partial_{\alpha}X^{\mu} \partial_{\beta}X^{\nu} \, , 
\end{equation}
where
\begin{subequations}\label{tau_H_GGK}
\begin{align}
        \tau_{\mu\nu} \dd X^{\mu} \dd X^{\nu} &= - \dd t^2 + \cos^2 t \dd z^2 \, , \\
        H_{\mu\nu} \dd X^{\mu} \dd X^{\nu} &= x^a x^a \left( - \dd t^2 + \cos^2 t \, \dd z^2 \right) + \dd x^a \dd x^a + \dd y^{a'} \dd y^{a'} \, , 
\end{align}
\end{subequations}
where $a \in \{ 1,2,3\}$, $a' \in \{ 1, ..., 5\}$ and indices are contracted with the Kronecker delta. Because $z$ is an isometry direction for the geometry \eqref{tau_H_GGK}, i.e. the geometry remains invariant under the shift $z \to z + k$ with $k= \text{const.}$, we can compactify the $z$ direction as 
\begin{equation}\label{compactification}
   z \simeq z + 2\pi \, . 
\end{equation}
Then, the static gauge fixing is achieved by imposing 
\begin{equation}\label{static_gauge}
    t = \tau \, , \qquad \qquad
    z = \sigma \, . 
\end{equation}
We remark that, thanks to the compactification \eqref{compactification}, closed string boundary conditions are compatible with the static gauge \eqref{static_gauge}. After fixing static gauge, the GGK action \eqref{GGK_NG} becomes,
\begin{equation}\label{GGK_static}
    S_{\text{GGK, bos}}^{\text{stat}} = - T \int \dd^2 \sigma \sqrt{|\tau|} \left[ 2 x^a x^a + \tau^{\alpha\beta} \left( \partial_{\alpha} x^a \partial_{\beta} x^a + \partial_{\alpha} y^{a'} \partial_{\beta} y^{a'} \right)\right] \, , 
\end{equation}
where $\tau_{\alpha\beta}$ is simply a metric of AdS$_2$ described in the worldsheet coordinates $(\tau, \sigma)$, given by 
\begin{equation}\label{AdS2_metric_static}
    \tau_{\alpha\beta} = \text{diag} (-1, \cos^2 \tau) \, . 
\end{equation}

We proceed by expanding the remaining eight degrees of freedom $(x^a,y^a)$ around a \emph{fully generic} classical closed string solution $(x^a_{\text{cl}},y^a_{\text{cl}})$, 
\begin{equation}\label{any_vacuum_expansion}
    x^a = x^a_{\text{cl}} + \frac{1}{\sqrt{T}} \, \delta x^a \, , \qquad\qquad
    y^{a'} = y^a_{\text{cl}} + \frac{1}{\sqrt{T}} \, \delta y^a \, .
\end{equation}
In the large string tension regime, i.e. $T \gg 1$, \eqref{GGK_static} expands around a generic vacuum solution with an \emph{exact} finite number of terms, namely 
\begin{equation}
    S^{\text{stat}}_{\text{GGK, bos}} = S_0 + S_1 + S_2 \, , 
\end{equation}
where 
\begin{subequations}
    \begin{align}
        S_0 &= - T \int \dd^2 \sigma \sqrt{|\tau|} \left[ 2 x^a_{\text{cl}} x^a_{\text{cl}} + \tau^{\alpha\beta} \left( \partial_{\alpha} x^a_{\text{cl}} \partial_{\beta} x^a_{\text{cl}} + \partial_{\alpha} y^{a'}_{\text{cl}} \partial_{\beta} y^{a'}_{\text{cl}} \right)\right] \, , \\
        S_1 &= - 2\sqrt{T} \int \dd^2 \sigma \sqrt{|\tau|} \left[ 2 x^a_{\text{cl}} \delta x^a + \tau^{\alpha\beta} \left( \partial_{\alpha} x^a_{\text{cl}} \partial_{\beta} \delta x^a + \partial_{\alpha} y^{a'}_{\text{cl}} \partial_{\beta} \delta y^{a'} \right)\right] \, , \\
        \label{S_2}
        S_2 &= - \int \dd^2 \sigma \sqrt{|\tau|} \left[ 2 \delta x^a \delta x^a + \tau^{\alpha\beta} \left( \partial_{\alpha} \delta x^a \partial_{\beta} \delta x^a + \partial_{\alpha} \delta y^{a'} \partial_{\beta} \delta y^{a'} \right)\right] \, .
    \end{align}
\end{subequations}
$S_0$ is simply a number once the classical solution is specified, and therefore, it can be disregarded. After integrating by parts, $S_1$ is proportional to the equations of motion for $x^a$ and $y^a$ evaluated on $(x^a_{\text{cl}},y^a_{\text{cl}})$, therefore vanishing up to a boundary term that is also cancelled by closed string boundary conditions.

In summary, the large tension expansion of GGK action in static gauge about \emph{any} vacuum is \emph{exact} at all orders, and captured by the quadratic action $S_2$ describing 3 massive fields $x^a$, with mass $m_B^2 = 2$, and 5 massless fields $y^{a'}$ in AdS$_2$.

%%%%%%%%%%%%%%%%%%%%%%%%%%%%%%%%%%%%%%%%%%%%%%%%%%%%%%%%%%%%%%%%%%%%%%%%%%%%%%%%%%%%%%%%%%%%%%%%%%%%%%%%%%%%%%%%%%%%%%%%%%%%%%%%%%%%%%%%%%%%%%%%%%%%%%%%%%%%%%%%%%%%%%%%%%%%%%%%%%%%%%%%%%%%%%%%%%%%%%%%%%%%%%%%%%%%%%%%%%%%%%%%%%%%%%%%%%%%%%%%%%%%%%%%%%%%%%%%%%%%%%%%%%%%%%%%%%%%%%%%%%%%%%%%%%%%%%%%%%%%%%%%%%%%%%%%%%%%%%%%%%%%%%%%%%%%%%%%%%%%%%%%%%%%%%%%%%%%%%%%%%%%%%%%%%%%%%%%%%%%%%%%%%%%%%%%%%%%%%%%%%%%%%%%%%%%%%%%%%%%

\subsection{The fermionic sector}

Let us move now to the fermionic sector of the theory. The non-relativistic limit of the full Green-Schwarz type IIB string action in AdS$_5\times$S$^5$ is discussed in \cite{Gomis:2005pg}, using the Metsaev-Tseytlin formulation \cite{Metsaev:1998it}. In 10D, spinors are described as two copies of a 16-components Majorana-Weyl spinor, $\theta^{I}$, with $I=1,2$.

First, we need to split the spinors $\theta^I$ by using the string Newton-Cartan projectors
\begin{equation}
    \theta^I_{\pm} \equiv P^{IJ}_{\pm} \theta^J \, , \qquad\qquad
    P^{IJ}_{\pm} \equiv \frac{1}{2}\left[\mathbbm{1} \pm \gamma^0 \gamma^1 \otimes  \mathbbm{1}_8 \otimes (\sigma_3)^{IJ}\right] \, , 
\end{equation}
where $\gamma^A = (\gamma^0, \gamma^1)$ are gamma matrices in 2D and $\sigma_3$ is the third Pauli matrix.  
After fixing static gauge as in \eqref{static_gauge}, and kappa-symmetry as follows\footnote{Notice that this gauge fixing is invariant under Galilei transformations of the target space. The most problematic of those to check are the Galilei boost, which mix a longitudinal and a transverse direction. Before taking the non-relativistic limit, the internal part of an AdS boost takes the schematic form $\delta_B \theta_{\pm} \to \theta_{\mp}$. To get this, we should use that Lie derivatives for spinors are written in terms of Kosmann derivatives $\mathsterling_{\xi} \theta = \xi^{\mu} \nabla_{\mu} \theta + \frac{1}{4} \nabla_{\mu} \xi_{\nu} \Gamma^{\mu\nu} \theta$. After taking the non-relativistic limit, which requires rescaling fermions as $\theta_+ \to \frac{1}{\sqrt{c}} \theta_+$ and $\theta_-\to \sqrt{c} \theta_-$, we obtain $\delta_G \theta_{\pm} \to c^{\pm 1}\theta_{\mp}$. Imposing the kappa-symmetry fixing \eqref{kappa_symmetry} eliminates the divergent contribution to $\theta_+$, while the $c\to \infty$ limit kills the contribution to $\theta_-$. Therefore, eq. \eqref{kappa_symmetry} is compatible with Galilei boost, i.e. $\delta_G \theta_- = 0$.}   
\begin{equation}\label{kappa_symmetry}
    \theta_-^I = 0 \, , 
\end{equation}
the full GGK action is described as 
\begin{equation}\label{GGK_full}
    S_{\text{GGK}}^{\text{stat}} = S_{\text{GGK, bos}}^{\text{stat}} + S_{\text{GGK, ferm}}^{\text{stat}} \, ,  
\end{equation}
where $S_{\text{GGK, bos}}^{\text{stat}}$ is given in \eqref{GGK_static}, and $S_{\text{GGK, ferm}}^{\text{stat}}$ is\footnote{We have normalised the fermionic fields to eliminate an overall factor of 2 compared with \cite{Gomis:2005pg}.} 
\begin{equation}\label{GGK_ferm}
  S_{\text{GGK, ferm}}^{\text{stat}} =  T \int \dd^2 \sigma \sqrt{|\tau|} \left[ i\delta_{IJ} \bar{\theta}^I_+ (\gamma^A \otimes \mathbbm{1}_8) \tau_A{}^{\alpha} \nabla_{\alpha} \theta^J_+  + \varepsilon_{IJ} \bar{\theta}_+^I \theta_+^J \right] \, .
\end{equation}
Here, $\varepsilon_{12} = +1$; $\tau_A{}^\alpha$ is the inverse vielbein of the AdS$_2$ metric \eqref{AdS2_metric_static}; $\bar{\theta}^I \equiv (\theta^{I})^{\dagger} \gamma^0 \otimes \mathbbm{1}_8$; $\nabla$ is the standard AdS$_2$ covariant derivative, which acts on spinors as
\begin{eqnarray}
    \nabla_{\alpha} \theta^I_+ = \left( \partial_{\alpha} + \frac{1}{4} \omega_{\alpha}{}^{AB} \gamma_{AB}\otimes \mathbbm{1}_8 \right)\theta^I_+ \, , 
\end{eqnarray}
where $\omega_{\alpha}{}^{AB}$ is the AdS$_2$ spin connection, whose only non-zero component is $\omega_{\alpha}{}^{01} = \dd \sigma \sin \tau$.  

To diagonalise the mass term, we introduce the complex spinors $\sqrt{2}\chi=\theta_+^1 + i \theta_+^2$ and $\sqrt{2}\psi= \theta_+^1 - i \theta_+^2$, and similarly for the bar fermions, giving us the following action
\begin{equation}\label{rot_GGK_ferm}
  S_{\text{GGK, ferm}}^{\text{stat}} =  iT \int \dd^2 \sigma \sqrt{|\tau|} \left[\bar{\chi} (\gamma^A \otimes \mathbbm{1}_8) \tau_A{}^{\alpha} \nabla_{\alpha} \chi  - \bar\chi \chi + \bar{\psi} (\gamma^A \otimes \mathbbm{1}_8) \tau_A{}^{\alpha} \nabla_{\alpha} \psi  +\bar\psi \psi   \right] \, .
\end{equation}
The minus sign difference in the mass term can be eliminated by performing the redefinition $\tilde{\chi} = (\gamma^0 \gamma^1) \chi$, which leaves the kinetic term invariant while changing the sign of the mass term. It is also worth emphasising that $\chi$ and $\psi$ are not independent spinors, one is the complex-conjugated of the other. With this change, and after combining the two spinors into $\Psi = (\tilde{\chi}, \psi)$, the fermionic action finally reads 
\begin{equation}\label{comb_GGK_ferm}
  S_{\text{GGK, ferm}}^{\text{stat}} =  iT \int \dd^2 \sigma \sqrt{|\tau|} \left[\bar{\Psi} (\gamma^A \otimes \mathbbm{1}_8) \tau_A{}^{\alpha} \nabla_{\alpha} \Psi  +\bar\Psi \Psi   \right] \, .
\end{equation}

When expanding the full action \eqref{GGK_full} around a generic classical solution, the classical value of any fermion is zero. Therefore the bosonic fields will continue to fluctuate as in \eqref{any_vacuum_expansion}, whereas the fermionic fields as 
\begin{equation}
    \Psi = 0 + \frac{1}{\sqrt{T}} \delta \Psi \, .
\end{equation}
In the large string tension expansion, the fermionic action \eqref{GGK_ferm} expands exactly as 
\begin{equation}\label{GGK_ferm_exp}
  S_{\text{GGK, ferm}}^{\text{stat}} =  i \int \dd^2 \sigma \sqrt{|\tau|} \left[\delta \bar{\Psi} (\gamma^A \otimes \mathbbm{1}_8) \tau_A{}^{\alpha} \nabla_{\alpha} \delta \Psi  +\delta \bar\Psi \delta \Psi   \right] \, .
\end{equation}
This action describes 8 Majorana fields in AdS$_2$, with mass $m_F = 1$. In the next section, we shall quantise the full GGK action, namely \eqref{S_2} plus \eqref{GGK_ferm_exp}, and find its exact spectrum around any vacuum.

%%%%%%%%%%%%%%%%%%%%%%%%%%%%%%%%%%%%%%%%%%%%%%%%%%%%%%%%%%%%%%%%%%%%%%%%%%%%%%%%%%%%%%%%%%%%%%%%%%%%%%%%%%%%%%%%%%%%%%%%%%%%%%%%%%%%%%%%%%%%%%%%%%%%%%%%%%%%%%%%%%%%%%%%%%%%%%%%%%%%%%%%%%%%%%%%%%%%%%%%%%%%%%%%%%%%%%%%%%%%%%%%%%%%%%%%%%%%%%%%%%%%%%%%%%%%%%%%%%%%%%%%%%%%%%%%%%%%%%%%%%%%%%%%%%%%%%%%%%%%%%%%%%%%%%%%%%%%%%%%%%%%%%%%%%%%%%%%%%%%%%%%%%%%%%%%%%%%%%%%%%%%%%%%%%%%%%%%%%%%%%%%%%%%%%%%%%%%%%%%%%%%%%%%%%%%%%%%%%%%

\subsection{The full spectrum}

Quantisation of free spin-0 and spin-1/2 fields in AdS$_2$ is a task that has already been done in the literature \cite{PhysRevD.18.3565, SAKAI1985661}. Therefore, we shall report the results, keeping the discussion brief. The first point to note, is that quatisation is typically done in a set of coordinates where the time coordinate is an isometry. This can be easily achieved via a worldsheet diffeomorphism, which takes the AdS$_2$ metric from \eqref{AdS2_metric_static} to    
\begin{equation}\label{AdS2_metric_time_iso}
    \tau_{\alpha\beta} = \frac{1}{\cos^2 \sigma} \text{diag} (- 1, 1) \, . 
\end{equation}
In this new set of coordinates, $\tau$ is an isometry direction, and $\sigma$ only appears through $\cos^2 \sigma$. Due to periodicity, we are allowed to compactify the target space embedding coordinate $\rho$ (that is identified with $\sigma$ in static gauge, $\rho = \sigma$) as $\rho \simeq \rho + 2\pi$. Physically, this new set of coordinates still describes a closed string.

Quantisation of the bosonic action \eqref{S_2} corresponds to quantising 3 massive $m_B^2 = 2$ and 5 massless scalar fields in AdS$_2$. The zero-point energy associated with the quantised oscillators produces the following contribution to the one-loop energy \cite{PhysRevD.18.3565,SAKAI1985661} (see also \cite{deLeeuw:2024uaq,Sakaguchi:2007ba} for the discussion of our specific case)
\begin{eqnarray}
    E_2^{\text{bose}} = \frac{1}{2} \sum_{n=0}^{\infty} \left[ 3 (n+2) + 5 (n+1) \right] \, . 
\end{eqnarray}
On the other hand, quantisation of the fermionic action \eqref{GGK_ferm_exp} corresponds to quantising 8 massive ($m_F = 1$) fermions in AdS$_2$. The zero-point energy of quantised oscillators gives the following one-loop energy contribution \cite{SAKAI1985661}
\begin{eqnarray}
    E_2^{\text{fermi}} = - \frac{1}{2}\sum_{n=0}^{\infty}  8 \left(n+\frac{3}{2} \right) \, . 
\end{eqnarray}
The total contribution is therefore 
\begin{eqnarray}\label{true_one-loop_energy}
    E_2 =  E_2^{\text{bose}} +  E_2^{\text{fermi}} =  - \frac{1}{2}\sum_{n=0}^{\infty} 1 \, , 
\end{eqnarray}
which is divergent. To remedy this, we consider instead the following regularised one-loop energy 
\begin{eqnarray}\label{regularised_one-loop_energy}
  E_2 (\epsilon) =  \frac{1}{2} \sum_{n=0}^{\infty} \left[ 3 (n+2) e^{-\epsilon (n+2)} + 5 (n+1) e^{-\epsilon (n+1)} - 8 \left(n+\frac{3}{2} \right) e^{-\epsilon (n+ 3/2)} \right] \, . 
\end{eqnarray}
Clearly, when $\epsilon=0$, we recover the initial series \eqref{true_one-loop_energy}. However, we can now first perform the series over $n$, which converges thanks to the exponential damping, and then expand it for a small $\epsilon$, and take $\epsilon \to 0$ at the very end. This is known as the \emph{Hurwitz series regularization}, where the zeta-function regularization is a special case of it.\footnote{The Hurwitz series is defined as follows, 
\begin{eqnarray*}
   \zeta(a,b) = \sum_{n=0}^{\infty} \frac{1}{(n+a)^b} \, ,\qquad  \mathfrak{Re} (b) > 1\, , \ a \neq 0, -1, -2, .... \, . 
\end{eqnarray*} 
Its analytic continuation is unique, and provided by the regularised series \eqref{Hurwitz_reg}.} Therefore, what we have is nothing else than a generalisation of the divergent spectrum that already affected the bosonic string theory, and  made finite via the zeta-function regularization $\zeta(-1) = - 1/12$ \cite{Gliozzi, Polchinski:1998rq}. This procedure leads us to the following unique regularization 
\begin{eqnarray}\label{Hurwitz_reg}
\sum_{n=0}^{\infty} (n+a) e^{-\epsilon (n+a)} = \frac{1}{\epsilon^2} + \frac{1}{12} (-1+6a-6a^2) + \mathcal{O}(\epsilon) \, . 
\end{eqnarray}
It turns out that the regularised one-loop energy \eqref{regularised_one-loop_energy} vanishes exactly, 
\begin{eqnarray}
    E_2(\epsilon) \qquad  \xrightarrow[\quad \epsilon\to 0 \quad]{} \qquad 
    \frac{1}{2\epsilon^2}\left( 3+5 - 8 \right)+ \frac{1}{2}\left( -\frac{13}{4} - \frac{5}{12} + \frac{11}{3} \right) = 0 \, .
\end{eqnarray}
This leads us to assert that the GGK theory, including fermions, has an exact spectrum about \emph{any} classical string vacuum that solely consists in the classical vacuum energy, with no loop contributions to it, 
\begin{eqnarray}
    E_{\text{GGK, any vacuum}} = E_{\text{cl}} \, . 
\end{eqnarray}
The effect of the non-relativistic limit is therefore of zooming near a BPS bound, in a way that any quantum correction disappears.

%%%%%%%%%%%%%%%%%%%%%%%%%%%%%%%%%%%%%%  Conclusions %%%%%%%%%%%%%%%%%%%%%%%%%%%%%%%%%%%%

\section{Conclusions}

In this article we solved the problem of finding the spectrum of the GGK theory at all loops and about any vacuum. We analysed this problem in two different gauges, the light-cone and the static gauges, finding results in agreement with each others. In light-cone gauge, we extended the work of \cite{deLeeuw:2024uaq} at all-loops, and we proved that the spectrum of the bosonic sector of GGK theory around the folded BMN-like and a number of other classical closed string solutions is always described by three massive and three massless free scalar fields in AdS$_2$. In static gauge, we showed a  similar result, but at an even stronger level: the GGK action expands about \emph{any} classical closed string vacuum in terms of three massive and five massless free scalars and eight massive Majorana fields in AdS$_2$. We computed the full spectrum by using the regularised Hurwitz series, and found that the quantum energy cancels out between bosonic and fermionic contributions, leaving us with a spectrum that is only characterised by the classical vacuum energy.

\paragraph{Holographic dual.}
The result of this article has an immediate application in the GGK/GYM holographic duality. The fact that the string spectrum of GGK theory around any vacuum has no contributions beyond the classical vacuum energy, strongly suggests that the holographic dual observable in the GYM theory with five scalar fields should gain no quantum corrections neither.\footnote{The GGK string spectrum would correspond in the dual theory to scaling dimensions of gauge invariant operators. This is a 
of the fact that a non-relativistic version of the state-operator correspondence can still be formulated in this case \cite{GYM_phase_space}.} It would be interesting to study the phase space of the GYM theory with five scalar fields and fermions, and see whether it turns out to be fully constrained, similarly to the case of its abelian version, Galilean Electrodynamics (GED) \cite{Hernandez:2026stx}.\footnote{On the other hand, the Feynman diagram structure of GYM in 2+1 dimensions was studied in \cite{Lambert:2026jtb}, suggesting the phase space is not fully constrained. Cancellations might occur once scalar fields and fermions are included, similarly to the GGK string spectrum.}  If future research confirms this scenario, then matching observables through the GGK/GYM dictionary will potentially boil down to a matter of matching classical quantities between the two theories.

\paragraph{Underlying integrability.} The fact GGK theory has an exact spectrum consisting in free fields in AdS$_2$ should be manifest from, or due to, an underlying integrability structure of the theory. Since the GGK theory is an example of String Newton-Cartan string theory, its integrability does not manifest with the same mechanism as in its parental relativistic AdS$_5\times$S$^5$ string theory. The coset action and its Lax pair were found in \cite{Fontanella:2022fjd, Fontanella:2022pbm} (see also \cite{Fontanella:2020eje} for a Lie algebra expansion approach), but its monodromy matrix becomes non-diagonalisable and its eigenvalues are independent of the spectral parameter \cite{Fontanella:2022wfj}. This leaves integrability in GGK theory an open challenge, see \cite{Fontanella:2026gaq} for a recent review. For other applications of integrability in String Newton-Cartan string theory, although different from the GGK theory, see \cite{Banerjee:2026lhz, Kluson:2017ufb, Roychowdhury:2019vzh}.  

\paragraph{Spectrum of the DLCQ string theory.} By applying T-duality on the longitudinal spatial isometry direction, GGK theory is mapped into type IIA string theory on a 5d time dependent pp-wave background times $\mathbb{R}^5$ \cite{Gomis:2005pg}, see also \cite{Bergshoeff:2018yvt, Gomis:2020izd} for a discussion of T-duality in non-relativistic string theory. Then, the energy spectrum of GGK theory is T-dual to the  Discrete Light-Cone Quantisation (DLCQ) spectrum of relativistic strings in a pp-wave background. The latter is not a standard task, due to the time dependency of the pp-wave background. It would be interesting to find the precise connection between the GGK spectrum and the DLCQ spectrum of type IIA string theory.

\paragraph{Spectrum of Spin Matrix theory.} In this article, we found the exact non-perturbative spectrum of GGK theory, namely the theory that appears after taking a non-relativistic limit on the target space of string theory in AdS$_5\times$S$^5$, while keeping  the worldsheet relativistic. However, on top of that, one could take a non-relativistic limit also on the worldsheet, leading to Spin Matrix theory \cite{Harmark:2017rpg}.\footnote{To make that possible, it seems a requirement that the GGK string must wind a longitudinal spatial direction that is an isometry. This is explicitly guaranteed in the GGK coordinates \eqref{tau_H_GGK}.}  The result obtained in this article seems to suggest that the spectrum of Spin-Matrix theory must also be made of the classical vacuum energy only, since an additional non-relativistic limit on the worldsheet should not introduce new quantum corrections to the energy. It would be interesting to clarify this point in the future.

\paragraph{Application to relativistic string theory.}
It would be interesting to explore which of the general classical solutions given by \eqref{gmodes} correspond to known relativistic solutions and perhaps find new ones. The method proposed in this article of  expanding the string action about a non-isometry direction with non-zero VEV could be used in other contexts as well. It would be interesting to see if this idea could turn useful to study the relativistic string action expanded around complicated string vacua.

%%%%%%%%%%%%%%%%%%%%%%%%%%%%%%%%%%%% Acknowledgments %%%%%%%%%%%%%%%%%%%%%%%%%%%%%%%%%%%% 

\section*{Acknowledgments}

We thank Federico Castellani for sparking the idea that led us to reconsider the static gauge. The only use of AI was ChatGPT suggesting the Hurwitz regulation of the spectrum. AF was partially supported by the Research Ireland Pathway Programme ``Integrability in Non-Lorentzian Holography'' (grant number 24/PATH-S/12557) and partially by the SFI-Royal Society (grant number RFF$\backslash$EREF$\backslash$210373). 
AF thanks Lia for her permanent support. MdL is supported by ERC-2022-CoG-FAIM 101088193.

%%%%%%%%%%%%%%%%%%%%%%%%%%%% APPENDICES %%%%%%%%%%%%%%%%%%%%%%%%%%%%%%%%%%%%%%%%%%%%

\setcounter{section}{0}
\appendix{Change of Variables}
\label{deriv}

Under the map \eqref{rhotosigma}, a generic field $\phi(\tau,\sigma)$ maps into
\begin{equation}
\Phi(\tau,\rho)=\phi(\tau,\sigma(\tau,\rho)) \,.
\end{equation}
Then, by the chain rule
\begin{equation}
    \phi_\tau \dd\tau+ \phi_\sigma \dd \sigma= \Phi_\tau \dd \tau +\Phi_\rho \dd \rho \,,
\end{equation}
and using the differential version of \eqref{rhotosigma},
\begin{equation}
    \dd \sigma=\left(1+\frac{\hbar}{\sqrt{2\kappa}} X_\rho\right)\dd\rho+\frac{\hbar}{\sqrt{2\kappa}} X_\tau \dd \tau \, , 
\end{equation}
we obtain the following equation between differentials, 
\begin{equation}
    \phi_\tau \dd \tau+\phi_\sigma \left(1+\frac{\hbar}{\sqrt{2\kappa}} X_\rho\right) \dd \rho + \phi_\sigma \frac{\hbar}{\sqrt{2\kappa}} X_\tau \dd \tau= \Phi_\tau \dd \tau +\Phi_\rho \dd \rho \,. 
\end{equation}
In order for this relation to hold, the derivatives of $\phi$ and $\Phi$ must fulfil,
\begin{equation}
   \phi_\sigma=\frac{1}{1+\frac{\hbar}{\sqrt{2\kappa}} X_\rho}\Phi_\rho, \qquad \phi_\tau=\Phi_\tau-\frac{\frac{\hbar}{\sqrt{2\kappa}} X_\tau}{1+\frac{\hbar}{\sqrt{2\kappa}} X_\rho}\Phi_\rho \,.
\end{equation}
For our particular problem, this reads as 
\begin{subequations}
    \begin{align}
        z_1' &=\frac{-\kappa \sec^2\left(\frac{\kappa \rho}{2}\right)}{1+\frac{\hbar}{\sqrt{2\kappa}} X_\rho}\, , & \dot{z_1}&=\frac{\sqrt{\frac{\kappa}{2}}X_\tau \hbar \sec^2\left(\frac{\kappa \rho}{2}\right)}{1+\frac{\hbar}{\sqrt{2\kappa}} X_\rho} \, , \\
        y_i' &=\frac{\hbar}{\sqrt{2\kappa}}\frac{Y_{i,\rho}}{1+\frac{\hbar}{\sqrt{2\kappa}} X_\rho}\, , & \dot{y_i} &=\frac{\hbar}{\sqrt{2\kappa}}Y_{i,\tau}-\frac{\frac{\hbar^2}{2\kappa} X_\tau}{1+\frac{\hbar}{\sqrt{2\kappa}} X_\rho}Y_{i,\rho} \, ,\\
        z_m'&=\frac{\hbar}{\sqrt{2\kappa}}\frac{(C_\rho Z_m+CZ_{m,\rho})}{1+\frac{\hbar}{\sqrt{2\kappa}} X_\rho} \, , & \dot{z}_m&=\frac{\hbar}{\sqrt{2\kappa}}CZ_{m,\tau}-\frac{\frac{\hbar^2}{2\kappa} X_\tau}{1+\frac{\hbar}{\sqrt{2\kappa}} X_\rho}(C_\rho Z_m+CZ_{m,\rho}) \, ,
    \end{align}
\end{subequations}
where, for convenience, we defined
\begin{equation}
    C=\cos(\kappa \rho) \sec^2\left(\frac{\kappa \rho}{2}\right) \,.
\end{equation}

%%%%%%%%%%%%%%%% BIBLIOGRAPHY %%%%%%%%%%%%%%%%

\bibliographystyle{nb}

\bibliography{Biblio.bib}

\end{document}